\documentclass[aps,nofootinbib,superscriptaddress,10pt]{revtex4}
\usepackage[utf8]{inputenc}
\usepackage[english]{babel}
\usepackage{graphicx}
\usepackage{graphicx}
\usepackage{amssymb,amsmath}

\begin{document}

\newcommand{\epp}{$e^+e^- \to \eta\pi^+\pi^-$ }
\newcommand{\fpi}{$e^+e^- \to \pi^+\pi^-\pi^0\pi^0$ }

\title{\boldmath
Measurement of the \epp cross section with the SND detector at the 
VEPP-2000 collider}

\begin{abstract}
The \epp cross section is measured at the SND detector in 
the $\eta$ decay mode $\eta\to 3\pi^0$. The analysis is based
on the data sample with an integrated luminosity of 32.7 pb$^{-1}$ collected
at the VEPP-2000 $e^+e^-$ collider in the center-of-mass energy range 
$\sqrt{s}=1.075-2.000$ GeV. The data obtained in the $\eta\to 3\pi^0$
decay mode are found to be in agreement with the previous SND
measurements in the $\eta\to \gamma\gamma$ mode. Therefore the
measurements in the two modes are combined.
\end{abstract}

\author{M.~N.~Achasov}
\author{A.~Yu.~Barnyakov}
\author{K.~I.~Beloborodov}
\author{A.~V.~Berdyugin}
\author{D.~E.~Berkaev}
\affiliation{Budker Institute of Nuclear Physics, SB RAS, Novosibirsk, 630090,
Russia}
\affiliation{Novosibirsk State University, Novosibirsk, 630090, Russia}
\author{A.~G.~Bogdanchikov}
\author{A.~A.~Botov}
\affiliation{Budker Institute of Nuclear Physics, SB RAS, Novosibirsk, 630090,
Russia}
\author{T.~V.~Dimova}
\author{V.~P.~Druzhinin}
\email[e-mail:]{druzhinin@inp.nsk.su}
\author{V.~B.~Golubev}
\author{L.~V.~Kardapoltsev}
\author{A.~G.~Kharlamov}
\affiliation{Budker Institute of Nuclear Physics, SB RAS, Novosibirsk, 630090,
Russia}
\affiliation{Novosibirsk State University, Novosibirsk, 630090, Russia}
\author{I.~A.~Koop}
\affiliation{Budker Institute of Nuclear Physics, SB RAS, Novosibirsk, 630090,
Russia}
\affiliation{Novosibirsk State University, Novosibirsk, 630090, Russia}
\affiliation{Novosibirsk State Technical University, Novosibirsk, 630092, Russia}
\author{A.~A.~Korol}
\affiliation{Budker Institute of Nuclear Physics, SB RAS, Novosibirsk, 630090,
Russia}
\affiliation{Novosibirsk State University, Novosibirsk, 630090, Russia}
\author{D.~P.~Kovrizhin}
\author{S.~V.~Koshuba}
\affiliation{Budker Institute of Nuclear Physics, SB RAS, Novosibirsk, 630090,
Russia}
\author{A.~S.~Kupich}
\affiliation{Budker Institute of Nuclear Physics, SB RAS, Novosibirsk, 630090,
Russia}
\affiliation{Novosibirsk State University, Novosibirsk, 630090, Russia}
\author{A.~P.~Lysenko}
\author{K.~A.~Martin}
\affiliation{Budker Institute of Nuclear Physics, SB RAS, Novosibirsk, 630090,
Russia}
\author{N.~A.~Melnikova}
\author{N.~Yu.~Muchnoi}
\affiliation{Budker Institute of Nuclear Physics, SB RAS, Novosibirsk, 630090,
Russia}
\affiliation{Novosibirsk State University, Novosibirsk, 630090, Russia}
\author{A.~E.~Obrazovsky}
\author{E.~V.~Pakhtusova}
\affiliation{Budker Institute of Nuclear Physics, SB RAS, Novosibirsk, 630090,
Russia}
\author{E.~A.~Perevedentsev}
\author{K.~V.~Pugachev}
\author{E.~V.~Rogozina}
\author{S.~I.~Serednyakov}
\author{Z.~K.~Silagadze}
\author{Yu.~M.~Shatunov}
\author{P.~Yu.~Shatunov}
\affiliation{Budker Institute of Nuclear Physics, SB RAS, Novosibirsk, 630090,
Russia}
\affiliation{Novosibirsk State University, Novosibirsk, 630090, Russia}
\author{D.~A.~Shtol}
\author{I.~K.~Surin}
\affiliation{Budker Institute of Nuclear Physics, SB RAS, Novosibirsk, 630090,
Russia}
\author{Yu.~A.~Tikhonov}
\affiliation{Budker Institute of Nuclear Physics, SB RAS, Novosibirsk, 630090,
Russia}
\affiliation{Novosibirsk State University, Novosibirsk, 630090, Russia}
\author{ Yu.~V.~Usov}
\affiliation{Budker Institute of Nuclear Physics, SB RAS, Novosibirsk, 630090,
Russia}
\author{A.~V.~Vasiljev}
\author{I.~M.~Zemlyansky}
\affiliation{Budker Institute of Nuclear Physics, SB RAS, Novosibirsk, 630090,
Russia}
\affiliation{Novosibirsk State University, Novosibirsk, 630090, Russia}

\maketitle

\section{Introduction}
In this paper we continue the study of the process \epp with the SND detector
at the VEPP-2000 $e^+e^-$ collider begun in Ref.~\cite{sndvepp2000}. 
This isovector process proceeding mainly via the $\rho\eta$ intermediate 
state~\cite{sndvepp2000} is important for spectroscopy of the excited 
$\rho$-like states, $\rho(1450)$ and $\rho(1700)$, and gives a sizable 
contribution into the total hadronic cross section at the center-of-mass 
(c.m.) energy region $\sqrt{s}=1.4-1.8$ GeV.  The \epp cross-section data 
can be used to predict the hadronic spectral function in the 
$\tau^-\to \eta\pi^-\pi^0\nu_\tau$ decay and thus to test the hypothesis of
conservation of vector current.

Previously the \epp process was studied in several 
experiments~\cite{nd,dm2,cmd2,babar,sndvepp2m,sndvepp2000}. The most complete
and accurate data were obtained by BABAR~\cite{babar} and SND at 
VEPP-2000~\cite{sndvepp2000}. In Ref.~\cite{sndvepp2000},  the $\eta$-meson was 
reconstructed via its decay mode $\eta\to 2\gamma$. The cross section was 
measured in the energy region from 1.22 to 2.00 GeV. Large background 
from the $e^+e^-\to \pi^+\pi^-\pi^0\pi^0$ and other hadronic processes 
didn't allow to perform measurement with comparable accuracy at lower energies.
In this work, we use the decay mode $\eta\to 3\pi^0$, in which the detection
efficiency is lower, but the signal-to-background ratio is better, to improve 
measurement sensitivity below 1.2 GeV.  

\section{Detector and experiment}
SND is a nonmagnetic detector~\cite{SND} collecting data at the VEPP-2000
$e^+e^-$ collider~\cite{VEPP} in the energy range $\sqrt{s}=0.3-2.0$ GeV. The 
direction and vertex position of charged particles are measured by a nine-layer
cylindrical drift chamber. Charged particle identification is based on $dE/dx$
measurements in the drift chamber and information from the system of threshold
aerogel Cherenkov counters. The photon energies and directions are measured in
a three-layer spherical electromagnetic calorimeter based on NaI(Tl) crystals.
The calorimeter covers a solid angle of about 95\% of $4\pi$. Its energy 
resolution for photons is 
$\sigma_{E_\gamma}/E_\gamma = 4.2\%/\sqrt[4]{E_\gamma({\rm GeV})}$, and
the angular resolution is about $1.5^\circ$. Outside the calorimeter, a muon
detector consisting of proportional tubes and scintillation counters is placed.

This work is based on a data sample with an integrated luminosity of 
32.7 pb$^{-1}$ collected in 2011-2012 in the c.m. energy range 
$\sqrt{s}=1.075-2$ GeV. The energy range was scanned several times with a step
of 25 MeV. During the experiment, the beam energy was determined using 
measurements of the magnetic field in the collider bending magnets. To fix the
absolute energy scale, the $\phi(1020)$ resonance mass measurement was 
performed. In 2012 the beam energy was measured in several energy points 
near 2 GeV by the back-scattering-laser-light system~\cite{COMPTON1,COMPTON2}.
The absolute energy measurements were used for calibration of the momentum
measurement in the CMD-3 detector, which collected data at VEPP-2000 
simultaneously with SND. The absolute c.m. energies for all scan points were
then determined using average momentum in Bhabha and $e^+e^-\to p\bar{p}$ 
events with accuracy of $2-6$ MeV~\cite{BEAM1}.

Simulation of the signal processes is done with the Monte Carlo (MC) event
generator based on formulas from Ref.~\cite{thepp} and uses the model of
the $\eta\rho(770)$ intermediate state. The generator takes into account 
radiative corrections to the initial particles calculated according to 
Ref.~\cite{radcor}. The angular distribution of additional photons radiated by
the initial particles is simulated according to Ref.~\cite{BM}. The \epp 
cross-section energy dependence needed for radiative-correction calculation is
taken from Ref.~\cite{sndvepp2000}. Interactions of the generated particles 
with the detector material are simulated using GEANT4 package~\cite{geant4}.
The simulation takes into account variation of experimental conditions during
data taking, in particular, dead detector channels and beam-induced background.
The beam background leads to appearance of spurious photons and charged 
particles in detected events. To take this effect into account, special 
background events recorded during data taking with a random trigger are used,
which are superimposed on simulated events.

The process of Bhabha scattering $e^+e^-\to e^+e^-$ is used for luminosity 
measurement. Accuracy of the luminosity measurement is estimated to be 
2\%~\cite{sndvepp2000}.

\section{Event selection}
\begin{figure}
\includegraphics[width=0.5\linewidth]{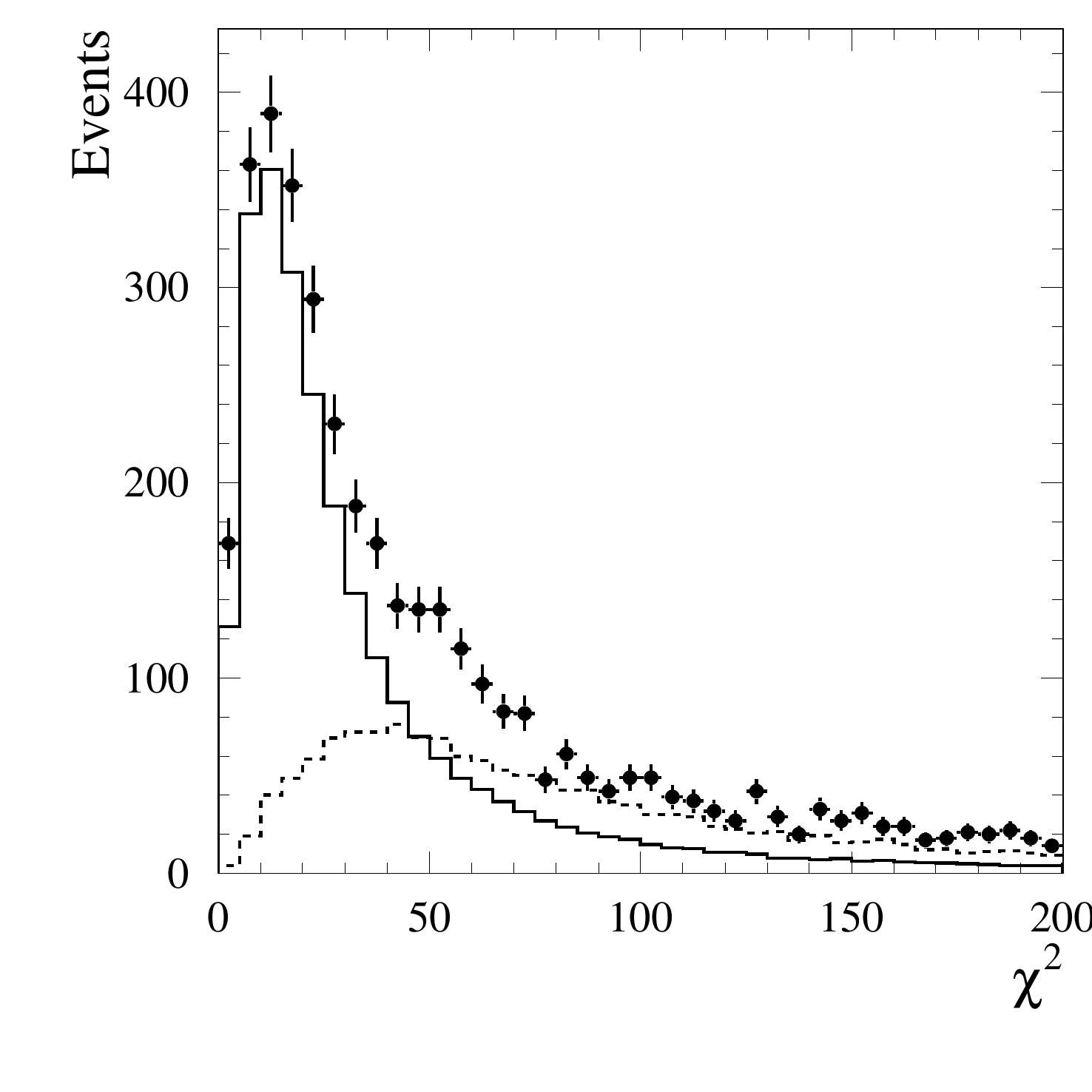}
\caption{The $\chi^2_{5\pi}$ distribution for data events with
$1.4\le \sqrt{s} \le 1.8$ GeV selected with the  additional condition
$500<M_{3\pi^0}<600$ MeV/$c^2$ (points with error bars) in comparison
with the simulated distributions for signal (solid histogram) and background
\fpi events (dashed histogram).
\label{fig:chi5pi}}
\end{figure}
In this analysis the $\eta$ meson is reconstructed via its decay 
$\eta\to 3\pi^0$. Therefore, we select events with two charged particles 
originated from the interaction region and at least six photons.

For selected events the vertex fit is performed using parameters of two 
charged tracks. The $\chi^2$ of the vertex fit ($\chi^2_{\rm vertex}$) is 
required to be less than 200. 
The found vertex is used to refine the parameters of charged particles and 
photons. Then the kinematic fit to the $e^+e^-\to \pi^+\pi^-3\pi^0$
hypothesis is performed with the requirement of energy and momentum balance 
and the $\pi^0$ mass constraints. The $\pi^0$ candidate is a two photon pair
with invariant mass in the range $90-200$ MeV/$c^2$. The quality of the 
kinematic fit is characterized by the parameter $\chi^2_{5\pi}$, which is 
required to be less than 45. If more than one photon combination satisfies
this condition, the combination with the smallest $\chi^2_{5\pi}$ value is 
chosen. Photon parameters corrected during the kinematic fit are used
to calculate the invariant mass of the three $\pi^0$ candidates ($M_{3\pi^0}$).

To suppress background from the process \fpi, the kinematic fit
to the \fpi hypothesis is performed, and the condition $\chi^2_{4\pi}>20$ is 
applied.

The $\chi^2_{5\pi}$ distribution for data events from the energy region 
$1.4\le \sqrt{s} \le 1.8$ GeV selected with the  additional condition  
$500<M_{3\pi^0}<600$ MeV/$c^2$ is shown in Fig.~\ref{fig:chi5pi} in 
comparison with the simulated distributions for signal and background
\fpi events.

\section{Determination of the number of signal events}
\begin{figure}
\includegraphics[width=0.5\textwidth]{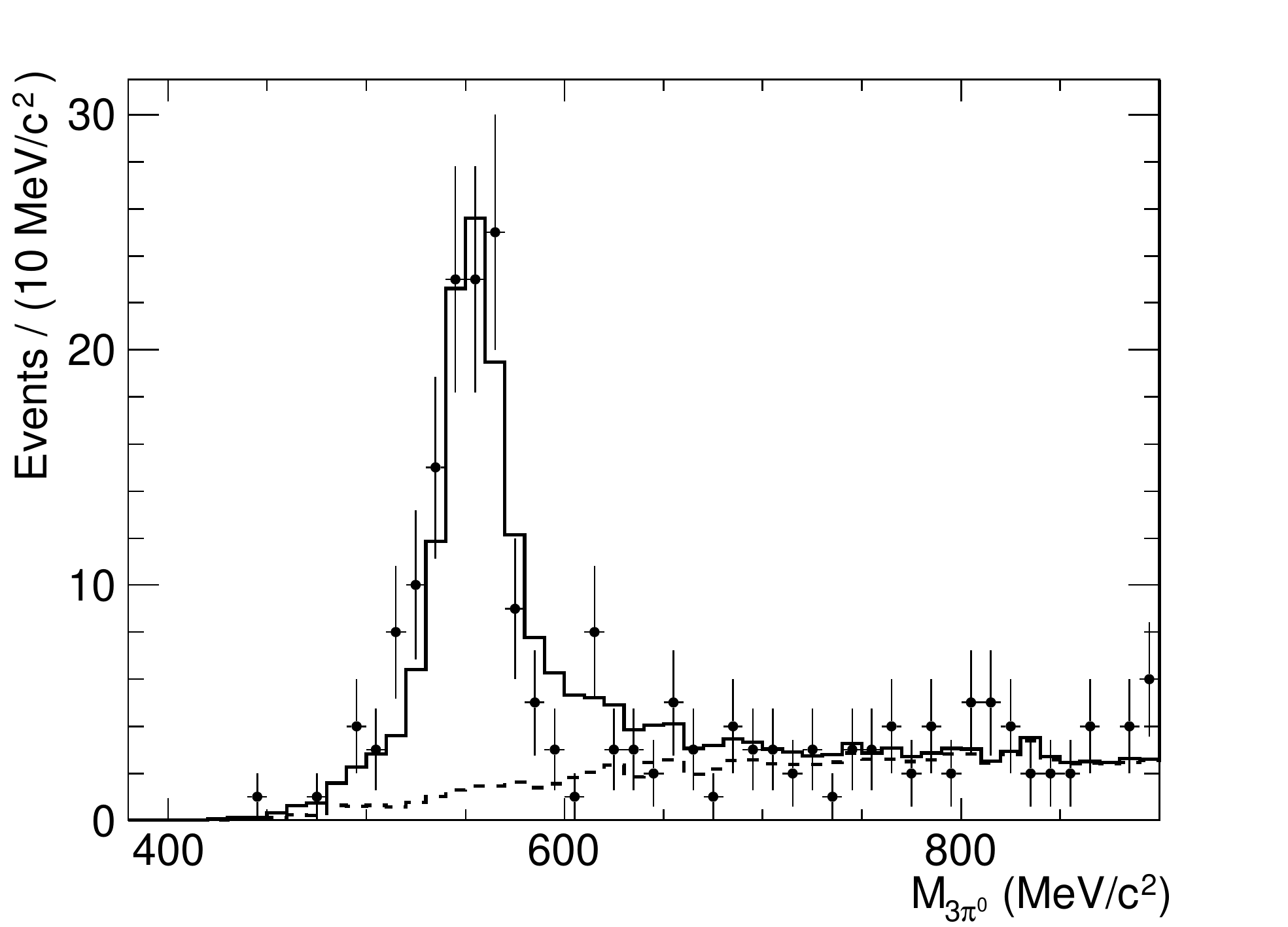} 
\caption{The $M_{3\pi^0}$ spectrum for selected data events with 
$\sqrt{s}=1.55$ GeV (points with error bars). The solid histogram is
the results of the fit by a sum of signal and background distributions. The
dashed histogram represents the fitted background spectrum. 
\label{fig2}}
\end{figure}
The $M_{3\pi^0}$ spectrum for selected data events with $\sqrt{s}=1.55$ GeV 
is shown in Fig.~\ref{fig2}. The spectrum is fitted with a sum of signal 
and background distributions. The signal distribution is described by a sum 
of three Gaussian functions with parameters determined from the fit to 
the $M_{3\pi^0}$ distribution for simulated signal events.
To account for a possible inaccuracy of the signal simulation, two parameters
are introduced: mass shift $\Delta M$ and a width correction $\Delta \sigma^2$.
The latter parameter is added to all Gaussian sigmas squared
($\sigma^2 = \sigma^2_{\rm MC}+\Delta \sigma^2$). These parameters are 
determined from the fit to the spectrum for data events from the energy 
interval near the maximum of the \epp cross section ($\sqrt{s}=1.45-1.60$ GeV)
and found to be $\Delta M=-(3.1 \pm 0.9)$ MeV/$c^2$ and 
$\Delta \sigma^2=-(24 \pm 19)$ MeV$^2$/$c^4$ for 2011 data set, and 
$-(3.4 \pm 1.5)$ MeV/$c^2$ and  $56 \pm 47$ MeV$^2$/$c^4$ for 2012 data set. 

\begin{figure}
\includegraphics[width=0.5\textwidth]{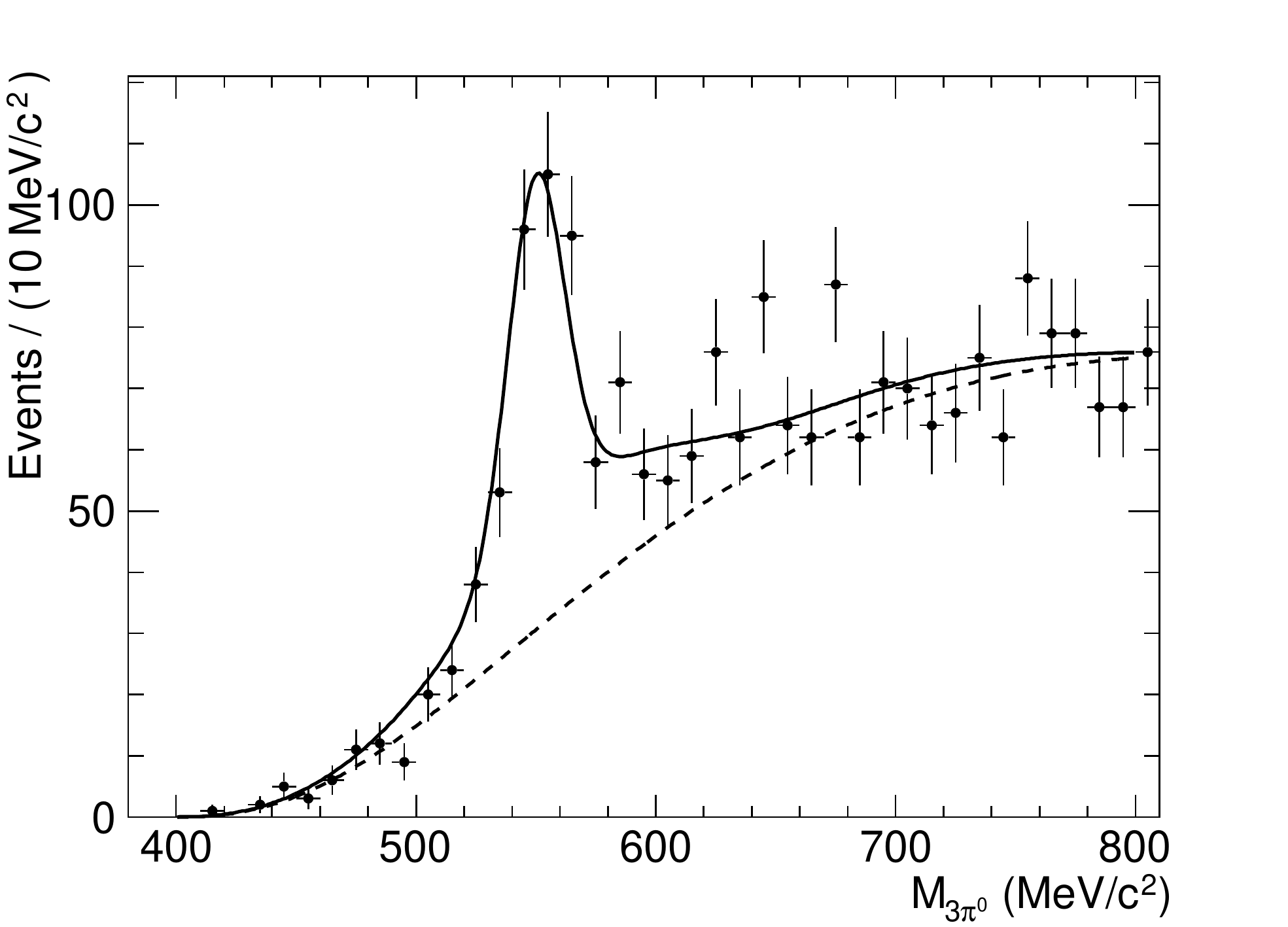}
\caption{The $M_{3\pi^0}$ spectrum for selected data events with 
$\sqrt{s}=1.7-2.0$ GeV (points with error bars). The solid curve is
the result of the fit by a sum of signal and background distributions. The
dashed curve represents the fitted background spectrum. 
\label{fig3}}
\end{figure}

The background distribution is obtained using simulation of the processes \fpi,
$e^+e^- \to \omega\pi^0\pi^0$, and $e^+e^- \to \pi^+\pi^-\pi^0\eta$. To 
calculate expected numbers of background events we use existing data on the
cross sections, in particular, the preliminary SND measurement~\cite{3pieta} 
for the $e^+e^- \to \pi^+\pi^-\pi^0\eta$ cross section.
A possible inaccuracy of background calculation is taken into account
by introducing a scale factor $\alpha_{\rm bkg}$. 
For energies below 1.6 GeV, the value of $\alpha_{\rm bkg}$ found in
the fit is consistent with unity. At higher energies, there is significant 
background contribution from other hadronic processes, e.g.,
$e^+e^-\to \pi^+\pi^-\pi^0\pi^0\eta$, or $e^+e^-\to \pi^+\pi^-4\pi^0$,
cross section for which are unknown.
In this region, the background is described by a function
based on the ARGUS distribution~\cite{argus}. It has been tested
that this function describes well the shape of the $M_{3\pi^0}$ spectra
for all background processes mentioned above. The example of the
fit with ARGUS background is shown in Fig.~\ref{fig3}.

To study the systematic uncertainty associated with the description
of background shape, the $M_{3\pi^0}$ spectrum for the energy region 
$\sqrt{s}=1.45-1.60$ GeV is fitted with the function based on the ARGUS 
distribution. The difference between numbers of signal events obtained
with this fit and the fit with the simulated background shape is found
to be 6\%. This number is taken as an estimate of the systematic 
uncertainty on the number of fitted signal events. 

The numbers of fitted \epp events for different energy points are listed
in Table~\ref{tab:bcs}.

\section{Detection efficiency\label{deteff}}
The detection efficiency is determined using MC simulation and then 
corrected for data-MC simulation difference in detector response:
$\varepsilon = \varepsilon_{MC}/(1-\Delta)$.
The correction for a specific selection criterion is calculated as
$\Delta = \frac{(N^\ast/N)_{\rm data}}{(N^\ast/N)_{\rm MC}}-1$,
where $N$ and $N^\ast$ are the numbers of signal events selected with 
the standard and loosened criterion.
\begin{table}
\caption{Efficiency corrections
\label{tab:delta}}
\begin{ruledtabular}
\begin{tabular}{lc}
Effect  & $\Delta$, \% \\
  \hline
  Condition $\chi^2_{\pi^+\pi^-3\pi^0}<45$ & $-1.5 \pm 2.7$ \\
  Condition $\chi^2_{\mathrm{vertex}}<200$     &  $0.9  \pm 0.4$ \\
  Track reconstruction & $0.3 \pm 0.2$ \\
  Photon conversion &  $2.0 \pm  0.2$\\
  \hline
  Total                 &  $1.7 \pm 2.8$\\
\end{tabular}
\end{ruledtabular}
\end{table}

The efficiency corrections are listed in Table~\ref{tab:delta}. 
To obtain the correction for the condition $\chi^2_{\pi^+\pi^-3\pi^0}<45$
we use data from the energy region $\sqrt{s}=1.425-1.68$ GeV and
change the boundary of the condition from 45 to 1000.
The corrections for the condition $\chi^2_{\rm vertex}<200$ and
track-reconstruction inefficiency are taken from Ref.~\cite{sndvepp2000}.
The data-MC simulation difference in photon conversion in detector material
before the tracking system is studied using events of the process 
$e^+e^-\to \gamma\gamma$.  

The corrected detection efficiency as a function of the c.m. energy is
listed in Table~\ref{tab:bcs}. Nonmonotonic behavior of the efficiency
is due to variations of experimental conditions (beam background, 
dead detector channels, etc.). The efficiency decrease above 1.6 GeV is 
explained by the decrease of the $e^+e^- \to \eta \pi^+\pi^-$ cross
section in this energy region and increase of the fraction of events 
with a hard photon radiated from the initial state, which are rejected
by the cut $\chi^2_{\pi^+\pi^-3\pi^0}<45$. 

The model dependence of the detection efficiency originating from
the uncertainty of the $e^+e^- \to \eta \pi^+\pi^-$
cross section used in simulation was studied in Ref.~\cite{sndvepp2000}.
It was found to be 1.0\% at $\sqrt{s}<1.6$ and 4.2\% GeV at higher energies. 

\section{\boldmath The \epp Born cross section}
\begin{table}
\scriptsize
\caption{\scriptsize The c.m. energy ($\sqrt{s}$), integrated luminosity ($L$),
number of signal events ($N$), detection efficiency ($\varepsilon$), 
radiative-correction factor ($1 + \delta$), \epp Born cross section 
measured in the  $\eta\to 3\pi^0$ decay mode ($\sigma_B$ for $\eta\to 3\pi^0$), 
and  \epp Born cross section combined with the SND measurement~\cite{sndvepp2000}
in the $\eta\to \gamma\gamma$ decay mode ($\sigma_B$). The quoted errors are
statistical. The systematic uncertainties are discussed in the text. For
the combined cross section it is 7\% below 1.45 GeV, 6\% at
$1.45<\sqrt{s}<1.6$ GeV, and 8\% above 1.6 GeV.
\label{tab:bcs}}
\begin{ruledtabular}
\begin{tabular}{ccccccc}
$\sqrt{s}$ (GeV) &  $L$ (nb$^{-1}$) & $N$ & $\varepsilon$ (\%) & $1+\delta$ &
$\sigma_B$ for $\eta\to 3\pi^0$ (nb) & $\sigma_B$ (nb)\\ 
\hline
1.075& 541& $ 11\pm  7$ & 6.0&0.874& $0.37\pm 0.25$ & $ 0.37\pm 0.25$ \\
1.097& 541& $  3\pm  4$ & 6.1&0.876& $0.09\pm 0.15$ & $ 0.09\pm 0.15$ \\
1.124& 528& $  4\pm  6$ & 6.1&0.877& $0.13\pm 0.20$ & $ 0.13\pm 0.20$ \\
1.151& 472& $  4\pm  4$ & 5.9&0.877& $0.17\pm 0.16$ & $ 0.17\pm 0.16$ \\
1.174& 532& $  2\pm  5$ & 6.0&0.876& $0.08\pm 0.19$ & $ 0.08\pm 0.19$ \\
1.196& 536& $  6\pm  4$ & 5.7&0.875& $0.21\pm 0.16$ & $ 0.21\pm 0.16$ \\
1.223& 553& $  8\pm  6$ & 5.8&0.873& $0.30\pm 0.23$ & $ 0.33\pm 0.13$ \\
1.245& 466& $  3\pm  4$ & 5.8&0.871& $0.13\pm 0.18$ & $ 0.15\pm 0.12$ \\
1.275&1225& $ 13\pm  8$ & 6.3&0.867& $0.19\pm 0.13$ & $ 0.33\pm 0.09$ \\
1.295& 484& $ 12\pm  5$ & 5.5&0.864& $0.53\pm 0.23$ & $ 0.51\pm 0.15$ \\
1.323& 542& $ 17\pm  6$ & 5.5&0.862& $0.67\pm 0.24$ & $ 0.71\pm 0.16$ \\
1.351&1398& $ 64\pm 11$ & 5.5&0.861& $0.96\pm 0.17$ & $ 1.02\pm 0.12$ \\
1.374& 599& $ 32\pm  8$ & 5.4&0.863& $1.17\pm 0.30$ & $ 1.22\pm 0.19$ \\
1.394& 643& $ 62\pm 10$ & 5.2&0.865& $2.13\pm 0.33$ & $ 1.86\pm 0.20$ \\
1.423& 591& $ 50\pm  9$ & 5.4&0.870& $1.79\pm 0.32$ & $ 2.06\pm 0.21$ \\
1.438&1442& $227\pm 18$ & 5.1&0.873& $3.54\pm 0.28$ & $ 3.03\pm 0.16$ \\
1.471& 608& $ 92\pm 12$ & 5.3&0.883& $3.21\pm 0.42$ & $ 3.29\pm 0.25$ \\
1.494& 731& $145\pm 14$ & 5.4&0.893& $4.10\pm 0.40$ & $ 3.82\pm 0.24$ \\
1.517&1395& $302\pm 21$ & 5.5&0.905& $4.35\pm 0.30$ & $ 4.44\pm 0.19$ \\
1.543& 566& $139\pm 14$ & 5.3&0.921& $5.07\pm 0.50$ & $ 4.55\pm 0.28$ \\
1.572& 436& $101\pm 12$ & 5.2&0.943& $4.69\pm 0.55$ & $ 3.94\pm 0.30$ \\
1.594& 446& $ 78\pm 15$ & 5.2&0.962& $3.54\pm 0.70$ & $ 3.34\pm 0.31$ \\
1.623& 530& $ 50\pm 13$ & 5.2&0.987& $1.85\pm 0.49$ & $ 3.12\pm 0.28$ \\
1.643& 490& $ 55\pm 15$ & 5.0&1.004& $2.25\pm 0.60$ & $ 2.45\pm 0.28$ \\
1.672&1314& $150\pm 22$ & 5.3&1.021& $2.09\pm 0.32$ & $ 2.30\pm 0.16$ \\
1.693& 472& $ 55\pm 13$ & 4.8&1.022& $2.36\pm 0.57$ & $ 2.67\pm 0.27$ \\
1.720&1022& $136\pm 17$ & 4.8&1.010& $2.75\pm 0.34$ & $ 2.23\pm 0.17$ \\
1.751&1197& $152\pm 24$ & 4.8&1.000& $2.63\pm 0.41$ & $ 2.36\pm 0.17$ \\
1.774& 473& $ 41\pm 11$ & 4.6&1.016& $1.84\pm 0.51$ & $ 1.96\pm 0.25$ \\
1.797&1391& $113\pm 22$ & 4.8&1.048& $1.61\pm 0.33$ & $ 2.00\pm 0.16$ \\
1.826& 513& $ 33\pm 10$ & 4.3&1.095& $1.38\pm 0.46$ & $ 1.44\pm 0.22$ \\
1.843&1369& $ 77\pm 19$ & 4.4&1.120& $1.14\pm 0.32$ & $ 1.31\pm 0.14$ \\
1.873&1556& $ 84\pm 18$ & 4.0&1.164& $1.16\pm 0.29$ & $ 0.97\pm 0.13$ \\
1.900&2033& $ 58\pm 21$ & 3.5&1.200& $0.69\pm 0.30$ & $ 0.79\pm 0.10$ \\
1.927&1256& $ 37\pm 14$ & 3.7&1.234& $0.66\pm 0.31$ & $ 0.76\pm 0.13$ \\
1.947&1312& $ 40\pm 15$ & 3.8&1.260& $0.63\pm 0.30$ & $ 0.71\pm 0.12$ \\
1.967& 724& $ 25\pm 12$ & 3.6&1.283& $0.75\pm 0.47$ & $ 0.74\pm 0.18$ \\
1.984&1125& $ 30\pm 15$ & 3.7&1.304& $0.56\pm 0.36$ & $ 0.74\pm 0.15$ \\
2.005& 576& $ 14\pm 11$ & 3.2&1.328& $0.57\pm 0.60$ & $ 0.78\pm 0.22$ \\
\end{tabular}
\end{ruledtabular}
\end{table}
The experimental value of the Born cross section for the $i$th energy
point is calculated as follows,
\begin{equation}
\sigma_{\mathrm{B},i}=\frac{N_i}{L_i(1+\delta_i)\varepsilon_i},
\label{eq1}
\end{equation}
where $L_i$ is the integrated luminosity, $N_i$ is the number of signal events,
$\varepsilon_i$ is the detection efficiency, and $\delta_i$ is the radiative
correction. The latter is determined as a result of the fit to data on the 
visible cross section
\begin{equation}
\sigma_{{\rm vis},i}=\frac{N_i}{L_i\varepsilon_i}
\end{equation}
with the function
\begin{equation}
\sigma_{\mathrm{vis}}(s)=
\int\limits_0^{z_{max}}\sigma_\mathrm{B}(s(1-z))F(z,s)dz=
\sigma_\mathrm{B}(s)(1+\delta(s)),
\label{eq2}
\end{equation}
where $F(z,s)$ is the function describing the probability of emission of 
photons with the energy $z\sqrt{s}/2$ by the initial electron and 
positron~\cite{radcor}, $z_{max}=1-(m_\eta+2m_\pi)^2/s$, and $m_\eta$ and
$m_\pi$ are the $\eta$ and $\pi^-$ masses.

The vector meson dominance (VMD) model with three intermediate isovector 
states, $\rho(770)$, $\rho(1450)$ and $\rho(1700)$, decaying into 
$\eta\rho(770)$~\cite{thepp} is used to describe the Born cross section: 
\begin{equation}
\sigma_\mathrm{B}(s) = \frac{4\pi\alpha^2}{3s^{3/2}}
\left | F_{\rho\eta\gamma}(s) \right |^2 P_f(s),
\end{equation}
where $\alpha$ is the fine structure constant, $F_{\rho\eta\gamma}(s)$
is the transition form factor for the vertex $\gamma^\ast\to \rho\eta$,
$P_f(s)$ is the function describing the energy dependence of
the $\eta\rho(770)$ phase space:
\begin{eqnarray}
P_f(s)&=&\frac{1}{\pi}\int_{4m_\pi^2}^{(\sqrt{s}-m_\eta)^2}
\frac{\sqrt{q^2}\Gamma_\rho(q^2)}
{(q^2-m^2_\rho)^2+q^2\Gamma^2_\rho(q^2)}
p^3(q^2) dq^2,
\label{eq5}\\
p(q^2)&=&\sqrt{\frac{(s-m^2_\eta-q^2)^2 - 4m^2_\eta q^2 }{4s}},
\nonumber\\
\Gamma_\rho(q^2)&=&\Gamma_\rho\frac{m^2_\rho}{q^2}
\left(\frac{q^2-4m_\pi^2}{m^2_\rho-4m_\pi^2} \right)^{\frac{3}{2}}.
\nonumber
\end{eqnarray}
where $m_\rho$ and $\Gamma_\rho$ are the $\rho(770)$ mass and width.
The transition form factor is parametrized as
\begin{equation}
F_{\rho\eta\gamma}(s)= 
\sum\limits_V g_Ve^{i\phi_V} \frac{m_V^2}
{s-m_V^2+i\sqrt{s}\Gamma_V(s)},\,\,\,V=\rho(770),\rho(1450),\rho(1700),
\label{eq:ff}   
\end{equation}
where $g_Ve^{i\phi_V}=g_{V\rho\eta}/g_{V\gamma}$
is the ratio is the coupling constants for the transitions 
$V\to\rho\eta$ and $V\to\gamma^\ast$.

The data on the visible cross section obtained in this work and in the previous
SND measurement in the $\eta\to \gamma\gamma$ decay mode~\cite{sndvepp2000} are
fitted simultaneously. The parameters of the $\rho(770)$ resonance are fixed at
the current world-average values~\cite{pdg}. The parameter $g_{\rho(770)}$
is calculated using the VMD relation 
$g_{\rho\rho\eta}=g_{\rho\gamma} g_{\rho\eta\gamma}$
from the $\rho(770)\to \eta\gamma$ decay width and is equal to 
$1.59\pm0.06$ GeV$^{-1}$. The phase $\phi_\rho(770)$ is set to zero.
\begin{figure}
\includegraphics[width=0.5\linewidth]{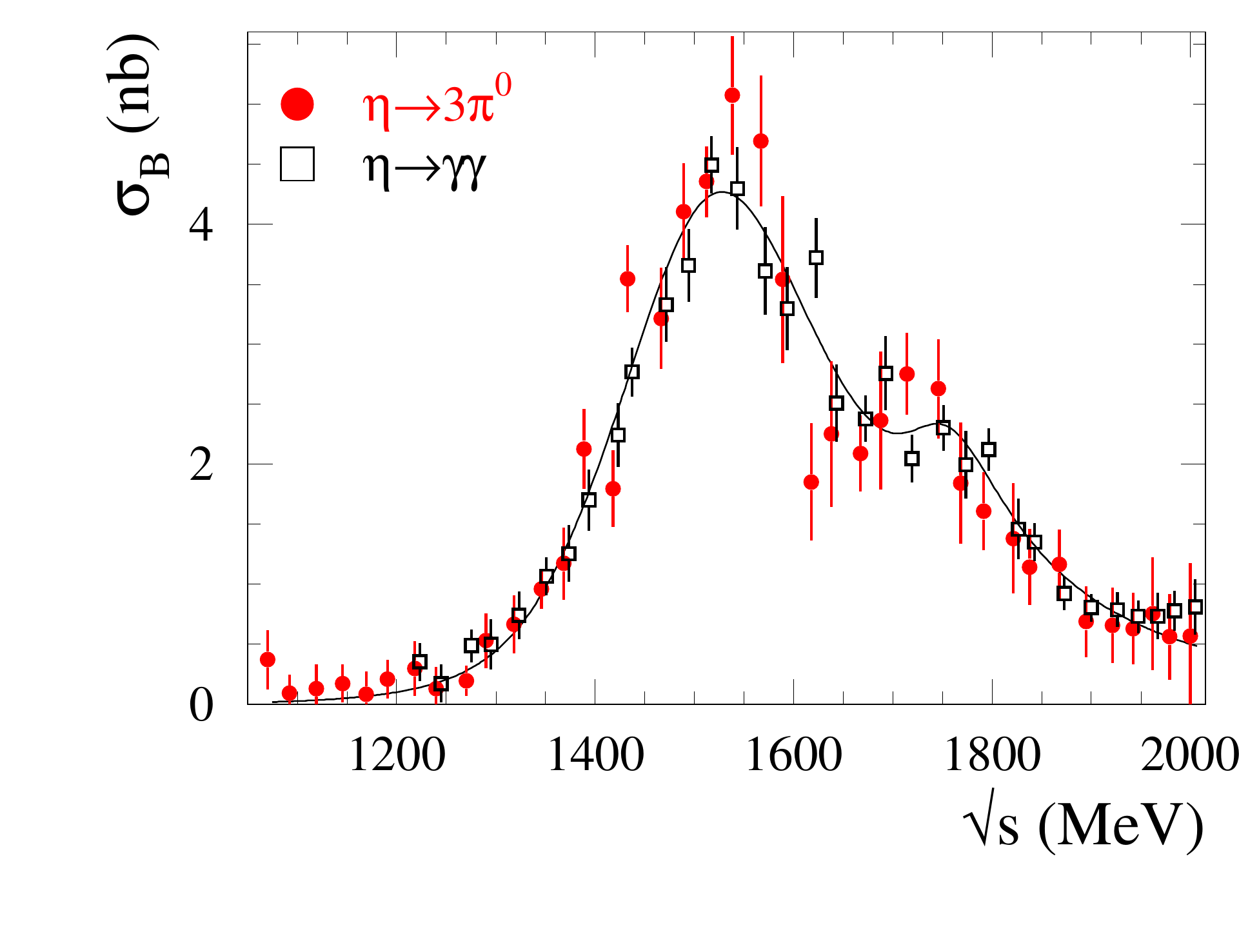}
\caption{The $e^+e^- \to \eta \pi^+\pi^-$ Born cross section 
measured by SND in the $\eta\to 3\pi^0$ and $\eta\to \gamma\gamma$ decay modes.
The curve is the result of the fit described in text.
\label{fig:cs0}}
\end{figure}

Following to Ref.~\cite{sndvepp2000} we assume that the coupling
constants $g_{V\rho\eta}$ and $g_{V\gamma}$ are real. So the phases
$\phi_\rho(1450)$ and $\phi_\rho(1700)$ can take values of 0 or $\pi$.
The masses, widths, and the constants $g_{\rho(1450)}$ and $g_{\rho(1700)}$ 
are free fit parameters.

The model with phases $\phi_\rho(1450)=\pi$ and $\phi_\rho(1700)=\pi$ describes
data well, $\chi^2/\nu=(37+31)/(39+33-6)=68/66$, where $\nu$ is the number
degrees of freedom. The first (second) numbers in the parentheses represent
the contribution from the data obtained in this work (Ref.~\cite{sndvepp2000}).
The values of the radiative correction calculated according to Eq.~(\ref{eq2}) 
and the values of the Born cross section obtained using Eq.~(\ref{eq1}) are 
listed in Table~\ref{tab:bcs}. The model uncertainty on the radiative 
correction is estimated by variation of the model parameters within their 
errors and is found to be 0.5\% below $\sqrt{s}=1.7$ GeV and 2\% above.
The systematic uncertainty on the cross section includes the systematic 
uncertainties on the number of signal events (6\%), detection efficiency (see 
Sec.~\ref{deteff}), radiative correction, and luminosity (2\%). It is equal
to 7\% below 1.6 GeV and 8\% above.

The comparison of the SND measurements
in the $\eta\to 3\pi^0$ and $\eta\to \gamma\gamma$ decay modes are 
presented in Fig.~\ref{fig:cs0}.
Since the data of the two measurements are consistent with each other, we
combine them. The combined cross section is listed in the last column of
the Table~\ref{tab:bcs}. For the first six energy points the measurement
are done only in the $\eta\to 3\pi^0$ mode. The systematic uncertainty on
the combined cross section is 7\% below 1.45 GeV, 6\% at
$1.45<\sqrt{s}<1.6$ GeV, and 8\% above 1.6 GeV.

\section{Discussion}
\begin{figure}
\includegraphics[width=0.5\linewidth]{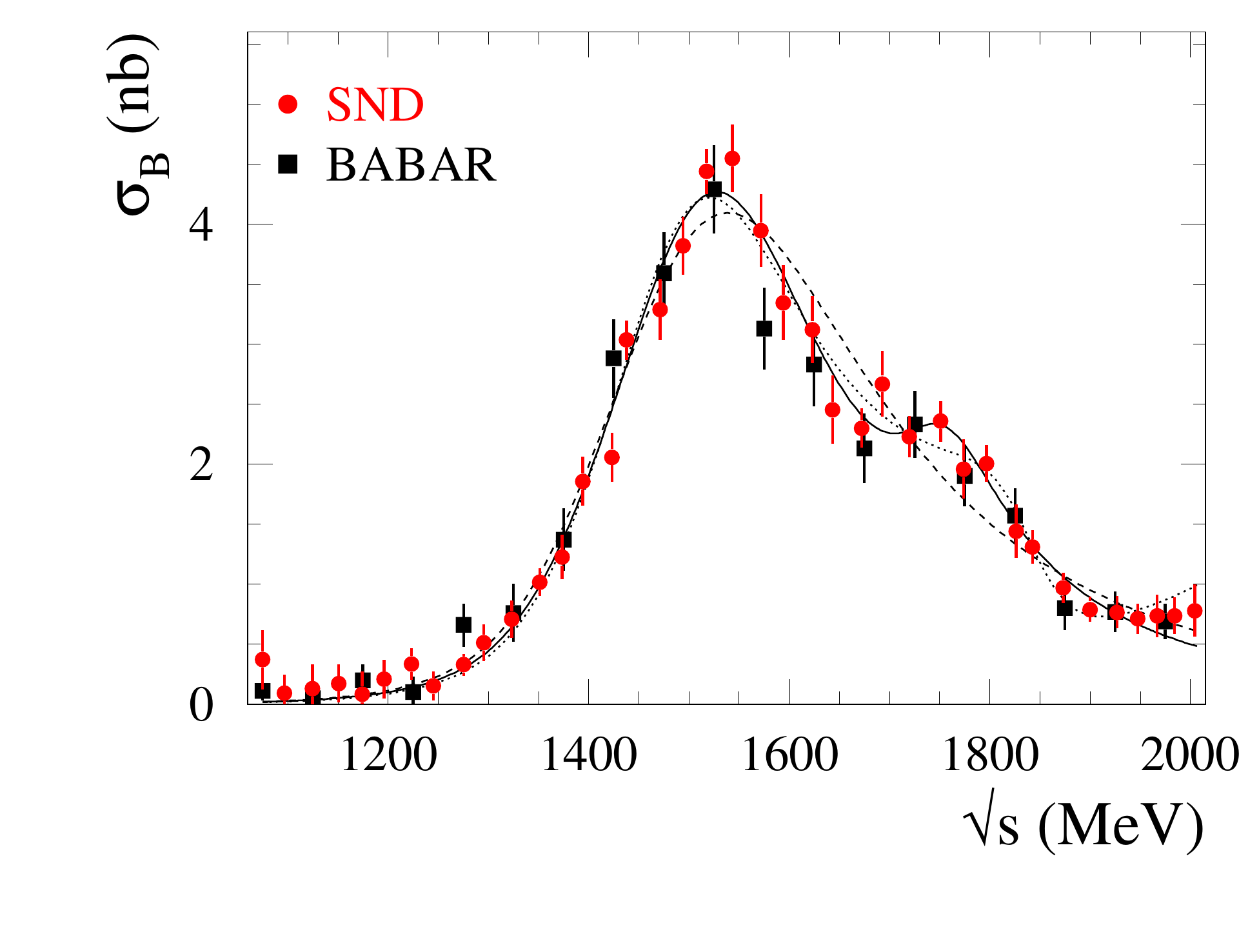}
\caption{The $e^+e^- \to \eta \pi^+\pi^-$ Born cross section measured by 
SND and BABAR~\cite{babar}. The solid, dashed, and dotted curves are
the results of the VMD fit with parameters listed in Table~\ref{tab:fit} 
for Models 1, 2, and 3, respectively.
\label{fig:cs1}}
\end{figure}
The comparison of the combined SND measurement with the previous most precise
data obtained by the BABAR Collaboration~\cite{babar} is presented in Fig.~\ref{fig:cs1}.
The two data sets are in agreement, but the SND data have better accuracy.
\begin{table}
\caption{Parameters of the VMD model.
\label{tab:fit}}
\begin{ruledtabular}
\begin{tabular}{lccc}
Parameter  & Model 1 & Model 2 & Model 3 \\
\hline
$g_{\rho(1450)}$ (GeV$^{-1}$)& $0.44\pm0.5$ & $0.56\pm0.2$ &$0.45\pm0.3$ \\ 
$\phi_{\rho(1450)}$ & $\pi$ & $\pi$ & $\pi$ \\
$m_{\rho(1450)}$ (MeV/$c^2$)& $1520\pm 10$ &$1510\pm 10$ & $1500\pm 10$ \\
$\Gamma_{\rho(1450)}$ (MeV)& $320\pm 30$ &$390\pm 10$ & $280\pm 20$ \\
$g_{\rho(1700)}$ (GeV$^{-1}$)& $0.024^{+0.019}_{-0.011}$ & $-$ &$0.025^{+0.014}_{-0.009}$ \\ 
$\phi_{\rho(1700)}$ & $\pi$ & $-$ & $0$ \\
$m_{\rho(1700)}$ (MeV/$c^2$)& $1750\pm 10$ &$-$ & $1840\pm 10$ \\
$\Gamma_{\rho(1700)}$ (MeV)& $135\pm 50$ &$-$ & $132\pm 40$ \\
$g_{\rho(2150)}$ (GeV$^{-1}$)& $-$ & $-$ & $0.084\pm0.008$ \\
$\chi^2/\nu$ & 33/33 & 55/36 & 29/32 \\
\end{tabular}
\end{ruledtabular}
\end{table}

The curves in Fig.~\ref{fig:cs1} represent the results of the fit to the SND 
data in the three models, which parameters are listed in
Table~\ref{tab:fit}. In all the models the phase $\phi_{\rho(1450)}=\pi$.
The fits with $\phi_{\rho(1450)}=0$ fail to describe data. Model 1 shown
by the solid curve is used in the previous section to calculate the radiative 
correction. It describes data well, but has a ``wrong'' value of 
$\phi_{\rho(1700)}$ equal to $\pi$. In the quark model~\cite{isgur} the 
$e^+e^-\to\rho(1450)\to\rho\eta $ and $e^+e^-\to\rho(1700)\to\rho\eta $ 
amplitudes are expected to be opposite in sign. The same prediction for 
the $e^+e^-\to\omega\pi$ process is confirmed in Ref.~\cite{ompi}.
The fit with the ``proper'' $\phi_{\rho(1700)}=0$ gives $g_{\rho(1700)}=0$
and coincides with Model 2 in Table~\ref{tab:fit}. This model shown in 
Fig.~\ref{fig:cs1} by the dashed curve describes data significantly worse, 
$P(\chi^2)=2\%$. It should be noted that 
in Ref.~\cite{sndvepp2000} Model 2 applied to the data obtained in the 
$\eta\to \gamma\gamma$ mode gave the reasonable value $P(\chi^2)=10\%$. So,
the addition of the new data obtained in the $\eta\to 3\pi^0$ mode 
strongly increases the significance of the $\rho(1700)$ signal. 

The reasonable quality of the fit with ``proper'' 
$\phi_{\rho(1700)}$ can be obtained in the model with an additional resonance
(Model 3 in Table~\ref{tab:fit}). The mass and width of this resonance are 
fixed at the PDG values $m_{\rho(2150)}=2155$ MeV/$c^2$ and 
$\Gamma_{\rho(2150)}=320$ MeV. The phase $\phi_{\rho(2150)}$ is set to zero.
The result of the fit is shown in Fig.~\ref{fig:cs1} by the dotted curve.
More precise data are needed to choose between Models 1 and 3. 

The parameters $g_V$ in the fit can be replaced by the products of
the branching fractions
\begin{equation}
B(V\to\rho\eta)B(V\to e^+e^-)=\frac{\alpha^2}{9}
\frac{g_V^2m_V}{\Gamma_V^2}P_f(m_V^2).
\end{equation}
The following values of the products are obtained
\begin{eqnarray}
B(\rho(1450)\to\rho\eta)B(\rho(1450)\to e^+e^-)\times10^7&=&(6.9\pm0.3)/(7.3\pm0.3),\\
B(\rho(1700)\to\rho\eta)B(\rho(1700)\to e^+e^-)\times10^8&=&(4.6^{+3.0}_{-1.9})/(8.3^{+3.8}_{-3.1})\nonumber
\end{eqnarray}
for Models 1 and 3, respectively.  
It is interesting that the parameters of the $\rho(1450)$ and $\rho(1700)$ 
resonances obtained in the two models with different relative phases of
the $\rho(1700)$ amplitude are rather close to each other. 

\section{Summary}
In this paper the cross section for the process \epp has been measured
in the c.m. energy range from 1.07 to 2.00 GeV in the decay mode 
$\eta\to 3\pi^0$. In the range 1.22--2.00 GeV the measured cross section
is found to be in good agreement with the previous SND measurement in the 
$\eta\to \gamma\gamma$ decay mode~\cite{sndvepp2000}. Therefore, the two 
measurements have been combined. 

The cross-section energy dependence has been fitted in the VMD model with 
2, 3 and 4 $\rho$-like states. The quality of the fit with two resonances,
$\rho(770)$ and $\rho(1450)$, is quite poor, $P(\chi^2)=2\%$, while
the fits with the additional $\rho(1700)$ resonance describe data well. 
The $\rho(1700)$ contribution appears as a shoulder on the $\rho(1450)$ peak
near 1.75 GeV. 

The SND data on the \epp cross section are in agreement with the previous most 
precise data obtained by the BABAR Collaboration~\cite{babar}, but have
better accuracy.

\section{ACKNOWLEDGMENTS}
Part of this work related to the photon reconstruction algorithm in the
electromagnetic calorimeter for multiphoton events is supported
by the Russian Science Foundation (project No. 14-50-00080).

\end{document}